\documentclass[%
 reprint,
 amsmath,amssymb,
 aps,prl
]{revtex4-1}

\usepackage{booktabs,array}
\usepackage{graphicx}
\usepackage{dcolumn}
\usepackage{bm}
\usepackage{amsmath,mathtools}

\begin{document}


\title{Towards a Resolution of the Static Correlation Problem in Density Functional Theory from Semidefinite Programming}

\author{Danny Gibney}
\author{Jan-Niklas Boyn}%
\author{David A. Mazziotti}%
\email{damazz@uchicago.edu}%
\affiliation{The James Franck Institute and The Department of Chemistry, The University of Chicago, Chicago, Illinois 60637 USA}%
\date{Submitted March 4, 2020}

\begin{abstract}
Kohn-Sham density functional theory (DFT) has long struggled with the accurate description of strongly correlated and open shell systems and improvements have been minor even in the newest hybrid functionals. In this Letter we treat the static correlation in DFT when frontier orbitals are degenerate by the means of using a semidefinite programming (SDP) approach to minimize the system energy as a function of the $N$-representable, non-idempotent 1-electron reduced density matrix. While showing greatly improved singlet-triplet gaps for linear density approximation and generalized gradient approximation (GGA) functionals, the SDP procedure reveals flaws in modern meta and hybrid GGA functionals, which show no major improvements when provided with an accurate electron density.
\end{abstract}

\maketitle


\textit{Introduction}-Since its conception in the 1960s, Kohn-Sham Density Functional Theory (KS-DFT) has become omnipresent in the calculation of the physical properties of atoms, molecules, liquids and solids. Its favorable computational scaling compared to wave function based methods such as coupled cluster or complete active space (CAS) calculations has made it into the foremost tool in computational catalysis and materials science~\cite{citationstats}. Nonetheless, current exchange correlation functionals continue to exhibit a long list of errors, including but not limited to the underestimation of chemical reaction barriers and band gaps of semiconductors, inaccurate description of spin state splittings and a general failure to describe systems with degenerate or near-degenerate electronic states~\cite{failures}. These failures can be traced to three fundamental issues: (i) a failure to capture accurately the long range $1/r^6$ asymptotic behavior of London dispersion forces~\cite{disp}; (ii) a delocalization error that arises in approximate functionals due to the dominating Coulomb term, leading to $\rho$ being artificially diffuse ~\cite{SIE, Perdew}; (iii) a static correlation error in near-degenerate states due to the single reference nature of KS-DFT~\cite{yang, constfrac}. \\

Over the course of the last five decades great progress has been made in functional development, significantly improving the accuracy of predictions for thermochemical properties, geometries, barrier heights, and other properties. Nonetheless, it has been shown that while modern functionals are trained to perform well in the calculation of specific chemical and physical properties, the error in computed electron densities has actually increased, suggesting the results may not reflect physical improvements to the functional but rather good fitting to the energy~\cite{medvedev,fitting}. Until DFT functionals can correctly account for both fractional spins and charges, a general solution will remain beyond reach and calculations will fail in systems where strong correlation plays a significant role. Progress on the static correlation error was presented in a recent letter by Lee and co-workers in which a new class of charge and spin densities is obtained via the breaking of time-reversal symmetry and the use of complex symmetry in the KS-DFT determinant, which they term ``complex polarization"~\cite{MHG}. \\

In this Letter we present a related and yet different approach to the static correlation problem in DFT that implements a semidefinite programming approach instead of the conventional KS self consistent field (SCF) procedure. Implemented with a boundary-point SDP algorithm~\cite{SDP2,SDP1} previously developed for variational 2-electron reduced density matrix (V2RDM) calculations~\cite{V22,V21,Schlimgen2016,Fosso-Tande2016,Mazziotti2016,Verstichel2012,Shenvi2010,Cances2006,Zhao2004,Mazziotti2002b,Nakata2001}, the SDP-DFT method variationally minimizes the energy with respect to the 1-electron reduced density matrix (1-RDM) subject to a set of $N$-representability constraints~\cite{Garrod1964,Coleman1963,Coleman2007,Mazziotti2012b} that ensure that the density represents physically viable system.  Unlike natural-orbital functional theory~\cite{Schmidt2019, Schilling2019, Piris2018, Piris2017a, Sharma2013, Rohr2008, Mazziotti2001b, Goedecker1998, BUIJSE2002}, the energy's correlation and exchange parts of the energy are evaluated with one-density exchange-correlation functionals from DFT.  Minimizing the energy with SDP allows idempotency breaking in the density matrix subject to the presence of degenerate frontier orbitals, yielding correct fractional-orbital spin densities, similar to those from complex spin-restricted orbitals, without the double-counting issue in wave-function-based multi-configurational DFT approaches~\cite{MCDFT1, MCDFT2, MCPDFT1, MCPDFT2}. We apply the SDP-DFT algorithm to calculate the singlet-triplet gaps of a set of 11 atoms and molecules surveying a range of commonly used DFT functionals.

\textit{Theory}-In contrast to the more complicated wave function based approaches, DFT uses the electron density as the basic variable of its calculations. The energy is minimized with respect to the electron density; the correct ground state density of a system is the one that minimizes the total energy through the functional $E[\rho(r)]$. This gives rise to the Kohn-Sham equations ~\cite{KSDFT}:
\begin{equation}
    E[\rho(r)]=T_s[\rho(r)] + E_{Ne}[\rho(r)] + J[\rho(r)] + E_{XC}[\rho(r)]\,,
\end{equation}
where $T_s$ is the kinetic energy functional, $E_{Ne}$ is the classical nuclear-electron Coulomb attraction, $J$ is the classical Coulomb repulsion and $E_{XC}$ is the exchange-correlation functional. The exact form of $E_{XC}[\rho(r)]$ remains unknown.
The electron density $\rho(r)$ may be computed from the 1-RDM $\prescript{1}{}{D}$:
\begin{equation}
    \rho(r) = 2\sum_{ij}\eta_i(r)\eta_j(r)\prescript{1}{}{D^i_j}\,,
\end{equation}
where $\eta$ are the molecular orbitals. In classical KS-DFT the 1-RDM is subject to trace, Hermicity and idempotency constraints. \\

Here we modify the traditional KS-DFT SCF approach to minimize the system energy over the convex set of $N$-representable 1-RDMs, an approach that has previously been demonstrated to obtain global solutions of restricted Hartree-Fock theory~\cite{SDPHF1, SDPHF2, Veeraraghavan2015}. The SDP replaces the diagonalization step in the SCF procedure, which in contrast to a traditional KS SCF implementation allows the 1-RDM to break idempotency and correctly account for orbital degeneracies. In the SDP the degenerate 1-RDM is obtained by minimizing  its trace against the 1-body reduced Hamiltonian, yielding the electronic energy:
\begin{equation}
   E_{GS} = \min \sum_{ij}\prescript{1}{}{H}^i_j\prescript{1}{}{D}^i_j \,,
\end{equation}
where $\prescript{1}{}{D}$ is subject to $N$-representability constraints~\cite{Garrod1964,Coleman1963,Coleman2007,Mazziotti2012b} to ensure that the density corresponds to a physically valid system. Namely, the $N$-representability constraints require $\prescript{1}{}{D}$ and the 1-hole matrix $\prescript{1}{}{Q}$ to remain positive semidefinite, meaning their eigenvalues remain nonnegative:
\begin{align}
    \prescript{1}{}{D} \succeq 0 \,, \\
    \prescript{1}{}{Q} \succeq 0  \,.
\end{align}
Additionally, the sum of $\prescript{1}{}{D}$ and $\prescript{1}{}{Q}$ needs to equal the one-particle identity matrix:
\begin{equation}
    \prescript{1}{}{D} + \prescript{1}{}{Q} = \prescript{1}{}{I}  \,.
\end{equation}
An additional constraint is levied on the trace of $\prescript{1}{}{D}$ which must equal the total number of electrons in the system:
\begin{equation}
    N = \sum_i \prescript{1}{}{D}^i_i    \,.
\end{equation}
Computationally this procedure is implemented using a semidefinite program in which a linear functional of matrices is minimized subject to linear constraints and the restriction that the matrices be positive semidefinite~\cite{SDP3,SDP4}.\\

As is clear from the Eqns. (4 - 6) no constraint is placed on the idempotency of $\prescript{1}{}{D}$, allowing us to obtain the correct non-idempotent 1-RDM with partial occupancies in the case of electronic degeneracies and, additionally, producing solutions that are eigenfunctions of the $\hat{S}^2$ and $\hat{S}_z$ operators. The SDP-DFT account for strong correlation from degenerate frontier orbitals that remains unaccounted for in the traditional single-reference KS-DFT approach.

\begin{table*}[t]
    \centering
    \caption{Root mean squared deviations (RMSDs) and mean signed deviations (MSDs) of the singlet-triplet gaps ($\Delta E_{\text{ST}} = E_\text{S} - E_\text{T}$) with respect to the experimental reference values for the test set of C, O, S, Si, NF, NH, O$_2$, PF, PH, S$_2$, SO. All values in kcal/mol.}
    \begin{ruledtabular}
    \begin{tabular}{ccccccccc}
    & \multicolumn{4}{c}{Restricted} & \multicolumn{4}{c}{Unrestricted}  \\
    \cmidrule(l){2-5}\cmidrule(l){6-9}
    & \multicolumn{2}{c}{SDP-DFT} & \multicolumn{2}{c}{KS-DFT} & \multicolumn{2}{c}{SDP-DFT} & \multicolumn{2}{c}{KS-DFT}  \\
    \cmidrule(l){2-3}\cmidrule(l){4-5}\cmidrule(l){6-7}\cmidrule(l){8-9}
    Functional  & RMSD & MSD & RMSD & MSD & RMSD & MSD & RMSD & MSD  \\
    \hline
    VWN & 5.33 & -3.83 & 17.85 & 17.56 & 4.68 & -3.03 & 14.56 & -13.86 \\
    SPW92 & 5.46 & -4.02  & 17.73 & 17.47 & 4.80 & -3.22 & 14.27 & -12.09\\
    PBE & 4.46 & -1.01  & 17.46 & 17.29 & 4.44 & 0.07  & 18.08 & -17.08  \\
    BLYP & 5.00 & -2.87  & 13.76 & 13.58 & 4.53 & -1.79 & 19.15 & -18.17 \\
    TPSS & 5.79 & 2.78 & 16.81 & 16.43 & 6.87 & 4.26 & 18.66 & -17.78 \\
    SCAN & 11.91 & 10.42 & 19.98 & 19.73 & 14.23 & 12.66  & 17.53 & -16.03 \\
    MN15-L & 10.11  & 8.70 & 9.78 & 9.17 & 12.92 & 11.47 & 12.34 & -11.21  \\
    B97M-V & 8.54 & 6.36 & 11.64 & 11.43 & 10.52 & 8.46 & 15.10 & -14.13 \\
    \end{tabular}
    \end{ruledtabular}
    \label{tab:Gaps}
\end{table*}

\textit{Applications}-We apply the SDP-DFT algorithm to a set of 11 atoms and molecules, previously assembled by Head-Gordon and co-workers for the purpose of benchmarking an electronic structure theory's ability to describe multi-reference character in atoms and small molecules from orbital degeneracies via the calculation of singlet-triplet energy gaps. We remove the charged species NO$^-$, leaving us a set consisting of C, O, S, Si, NF, NH, O$_2$, PF, PH, S$_2$, SO. In line with the previous studies on this data set, we are using a range of functionals covering the different rungs of Jacob's Ladder~\cite{Ladder}, namely traditional local density approximation (LDA) functionals VWN~\cite{LDA1, LDA2, Slater, VWN} and SPW92~\cite{SPW}, generalized gradient approximation (GGA) functionals PBE~\cite{PBE1,PBE2} and BLYP~\cite{BLYP1,BLYP2}, and some of the latest meta-GGA functionals TPSS~\cite{TPSS}, SCAN~\cite{SCAN}, MN15-L~\cite{MN15L}, and B97M-V~\cite{B97MV}. The augmented correlation-consistent polarized valence quadruple-zeta (aug-cc-pVQZ) basis set~\cite{basis1, basis2} was used for all calculations. \\

To quantify the performance of DFT functionals within the SDP framework we calculate the singlet-triplet gaps ($\Delta E_{\text{ST}} = E_{\text{S}} - E_{\text{T}}$) for the 11 entities in our test set with the chosen functionals and compare those against reference KS-DFT calculations and reported experimental values. Statistical analysis of the performance of the various functionals in spin restricted and spin unrestricted implementations is provided in Table I in the form of root mean squared deviation (RMSD) and mean signed deviation (MSD) from the experimental reference values. Equivalent results from KS-DFT calculations are provided for comparison. \\

In traditional unrestricted KS-DFT (UKS-DFT) all functionals significantly underestimate $\Delta E_{\text{ST}}$. The newer Minnesota functional MN15-L gives the best performance with a RMSD of 12.34 kcal/mol and a MSD of -11.21 kcal/mol, while BLYP performs worst with a RMSD of 19.15 kcal/mol and a MSD of -18.17 kcal/mol. These are large errors compared to chemical accuracy of 1 kcal/mol and the improvement of the newest meta-GGA functionals over the 40-year-old VWN is minor at best. Use of the spin unrestricted SDP-DFT (USDP-DFT) algorithm yields greatly improved results over UKS-DFT, with the largest increases in accuracy observed for the tested LDA and GGA functionals, all of which give RMSDs in the 4-5 kcal/mol range. The popular PBE functional performs best with the RMSD reduced to 4.44 kcal/mol. Of the tested meta-GGA functionals only TPSS shows a sizeable improvement with an RMSD of 6.87 kcal/mol using USDP-DFT, compared to 18.66 kcal/mol in UKS-DFT. MN15-L, the best performing functional in UKS-DFT, is the only functional tested to perform worse in USDP-DFT, albeit only slightly with an increase in RMSD of 0.58 kcal/mol. \\

Applying restricted SDP-DFT (RSDP-DFT) yields results comparable to the unrestricted calculations. While the RMSD is similar for the various functionals across restricted and unrestricted KS-DFT we observe a sign change in the MSD and $\Delta E_{\text{ST}}$ is significantly overestimated rather than underestimated in a restricted calculation. Again MN15-L gives the best performance compared to experiment. In contrast to this, the SDP results for restricted calculations mirror those obtained via an unrestricted implementation. Compared to USDP-DFT, the LDA and GGA functionals yield slightly increased RMSDs and MSDs in RSDP-DFT. The meta-GGA functionals show small improvements. MN15-L again performs worse with the SDP algorithm than without it.  \\

\begin{table}[b]
    \centering
    \caption{Mean energy differences of total electronic energy between the SDP-DFT and KS-DFT solutions ($\Delta E = (E_{\text{SDP}} - E_{\text{DFT}})/N$) of singlet and triplet states in both spin restricted and unrestricted frameworks in kcal/mol.}
    \begin{ruledtabular}
    \begin{tabular}{ccccc}
    & \multicolumn{2}{c}{Restricted} & \multicolumn{2}{c}{Unrestricted} \\
    \cmidrule(l){2-3}\cmidrule(l){4-5}
    Functional  & $\Delta E^\text{S}_{\text{tot}}$ & $\Delta E^\text{T}_{\text{tot}}$ & $\Delta E^\text{S}_{\text{tot}}$ & $\Delta E^\text{T}_{\text{tot}}$ \\
    \hline
     VWN &  -21.60 & -0.21 & 11.26 &  0.20  \\
     SPW92 & -21.62 & -0.12 & 11.02 & 0.19 \\
     PBE  & -17.31 & 0.99 & 18.75 & 1.60\\
     BLYP & -15.45 & 1.00 & 17.80 & 1.42 \\
     TPSS & -10.97 & 2.69 & 25.14 & 3.10 \\
     SCAN & -5.26  & 4.05 & 33.34 & 4.65  \\
     MN15-L & 4.42 & 4.89 & 27.88 & 5.20\\
     B97M-V & -1.08 & 3.99 & 26.77 & 4.18\\
    \end{tabular}
    \end{ruledtabular}
    \label{tab:Delta}
\end{table}

Average changes of the total electronic energy between SDP-DFT and KS-DFT for the singlet and triplet spin states of the tested functionals in both spin restricted and unrestricted formalisms are shown in Table II. In both restricted and unrestricted calculations the changes to the electronic energy of the triplet states are minor. LDA functionals VWN and SPW92 show minimal variation in $\Delta E_{\text{tot}}$ ($\approx 0.4$~kcal/mol) upon changing from a KS to a SDP algorithm while all other functionals show a small energy increase ranging from 0.99 to 4.89 kcal/mol in a restricted and 1.60 to 5.20 kcal/mol in an unrestricted framework. It is worth noting that these changes are entirely driven by changes from the atoms in the data set and both the total energies as well as their individual components remain unchanged upon use of the SDP in the molecules of the data set. This suggests an overestimation of the correlation energy in the triplet state of the atoms in the data set by GGA and meta-GGA functionals in traditional KS-DFT. \\

Variations are of significantly greater magnitude in the singlet state. In a spin restricted formalism the singlet state is lowered by the SDP optimization in all functionals but MN-15L, with the greatest energetic gains in LDA functionals, $\Delta E^S_{\text{tot}} = -21.62$ kcal/mol in SPW92, and successive decreases as we progress up Jacob's Ladder to $\Delta E^S_{\text{tot}} = 4.42$ kcal/mol in MN15-L. In a spin unrestricted formalism singlet states are raised in energy and the functionals follow the opposite trend to the restricted formalism, with the smallest change in LDA functionals and the greatest in meta-GGA functionals, ranging from $\Delta E^S_{\text{tot}} = 11.02$ kcal/mol in SPW92 to $\Delta E^S_{\text{tot}} = 33.34$ kcal/mol in SCAN. The lowering in energy upon introduction of the non-idempotent, correlated density via the RSDP-DFT procedure is expected, as the static correlation energy of the open shell singlet is recovered. Conversely, in USDP-DFT , raising of the singlet energy by introducing the correlated density points towards over-correlation of the open shell singlet by symmetry breaking in the UKS-DFT. The lack of energetic gain in USDP-DFT of meta-GGA functionals and their strong overcompensation in RKS-DFT suggests that the increasingly parameterized nature of these functionals seems to have resulted in a non-physical relationship between density and correlation energy, i.e. they have been fitted to predict the idempotent density matrix to yield a lower energy than the non-idempotent density matrix corresponding to a strongly correlated system. \\

Across all calculations the use of the non-idempotent SDP 1-RDM to evaluate the system energy leads to uniform changes in the individual components: a decrease in $E_{\text{kin}}$ and $E_{\text{XC}}$ counteracted to a varying degree by increases in $E_{\text{C}}$ and $E_{\text{nuc}}$. Additional details can be found in Table S1 of the Supplemental Information. Inspection of the energy components reveals very little consistency in how the different functionals react to the introduction of the non-idempotent 1-RDM. Only the simplest LDA functionals VWN and SPW92 react identically to the SDP density, while the more highly parameterized GGA and meta-GGA functionals show highly inconsistent changes in their individual components of the total electronic energy, emphasizing the strongly, often empirically, fitted nature of these functionals. \\

\begin{figure*}
    \begin{minipage}[b]{0.49\textwidth}
    \centering
    \includegraphics[scale=0.30]{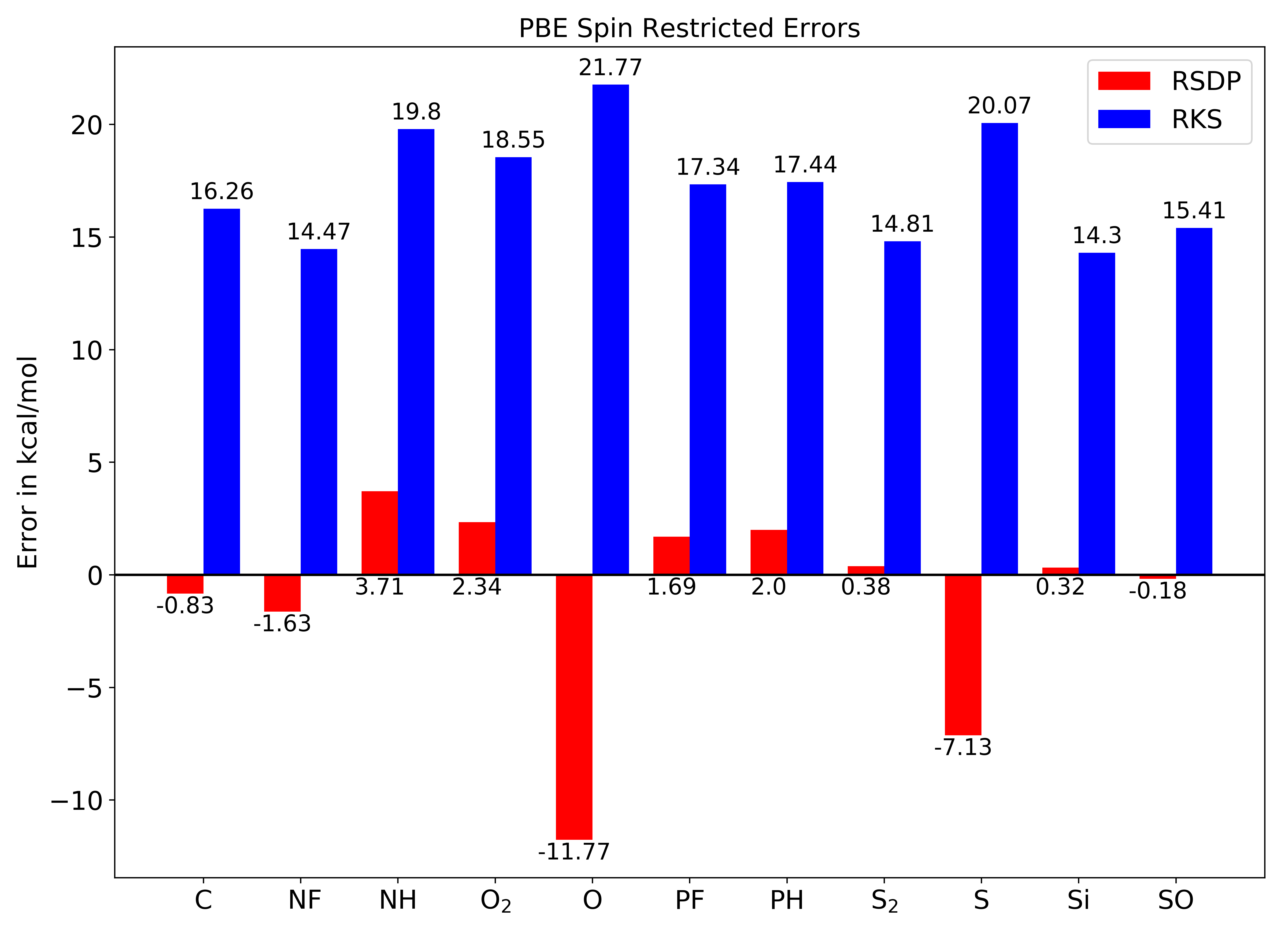} \\
    \includegraphics[scale=0.30]{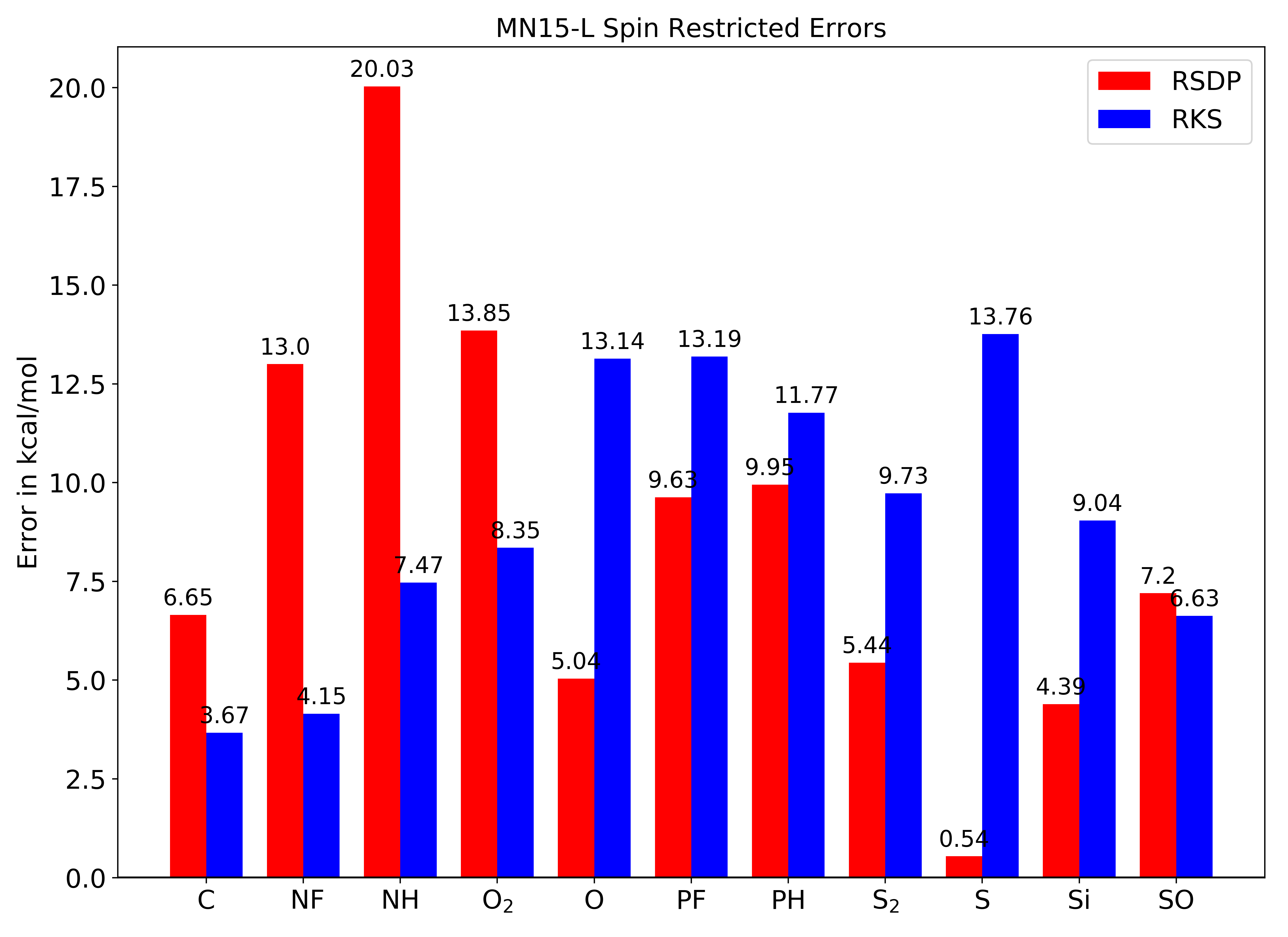}
    \end{minipage}
    \begin{minipage}[b]{0.49\textwidth}
    \includegraphics[scale=0.30]{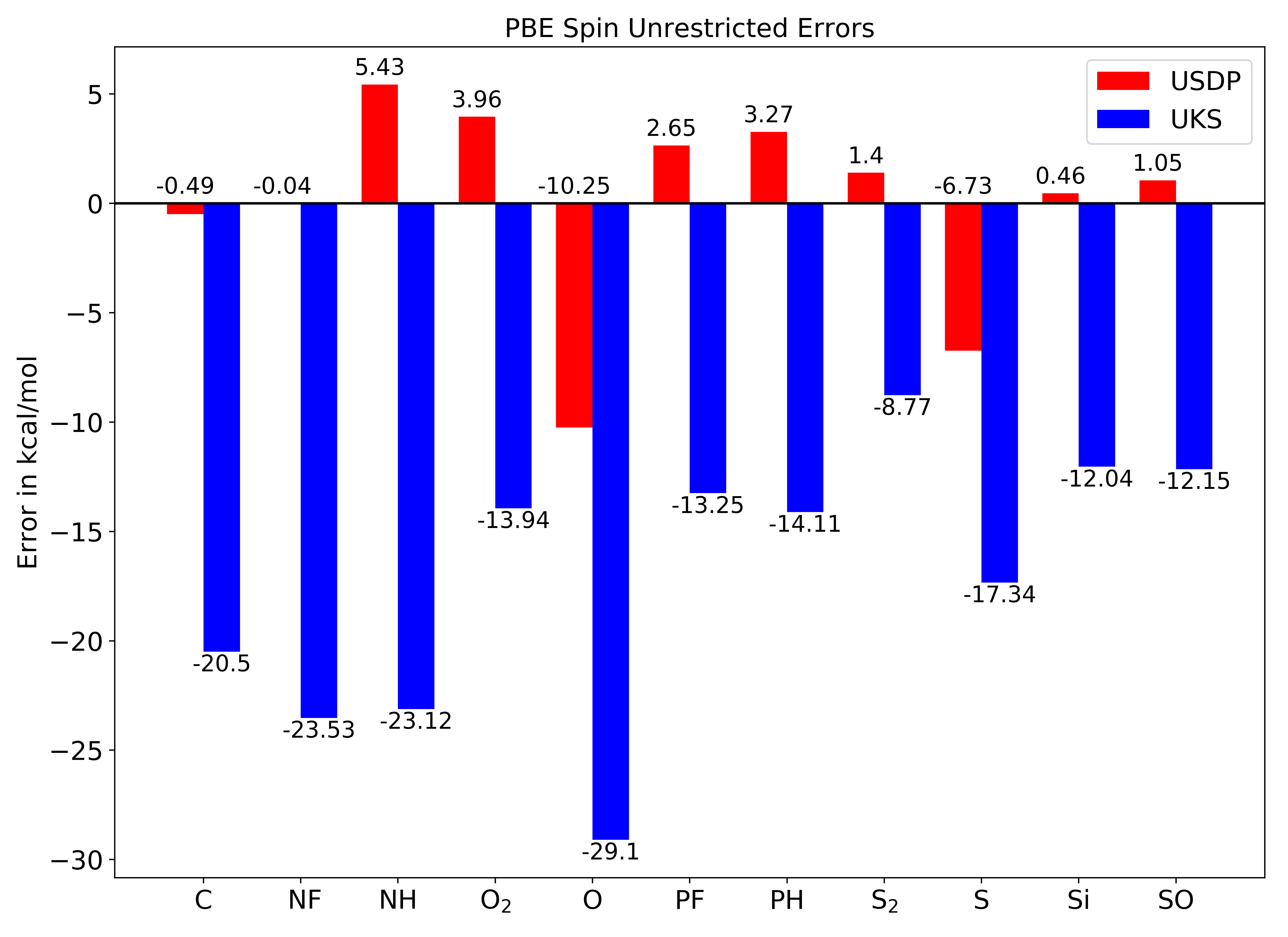} \\
    \includegraphics[scale=0.30]{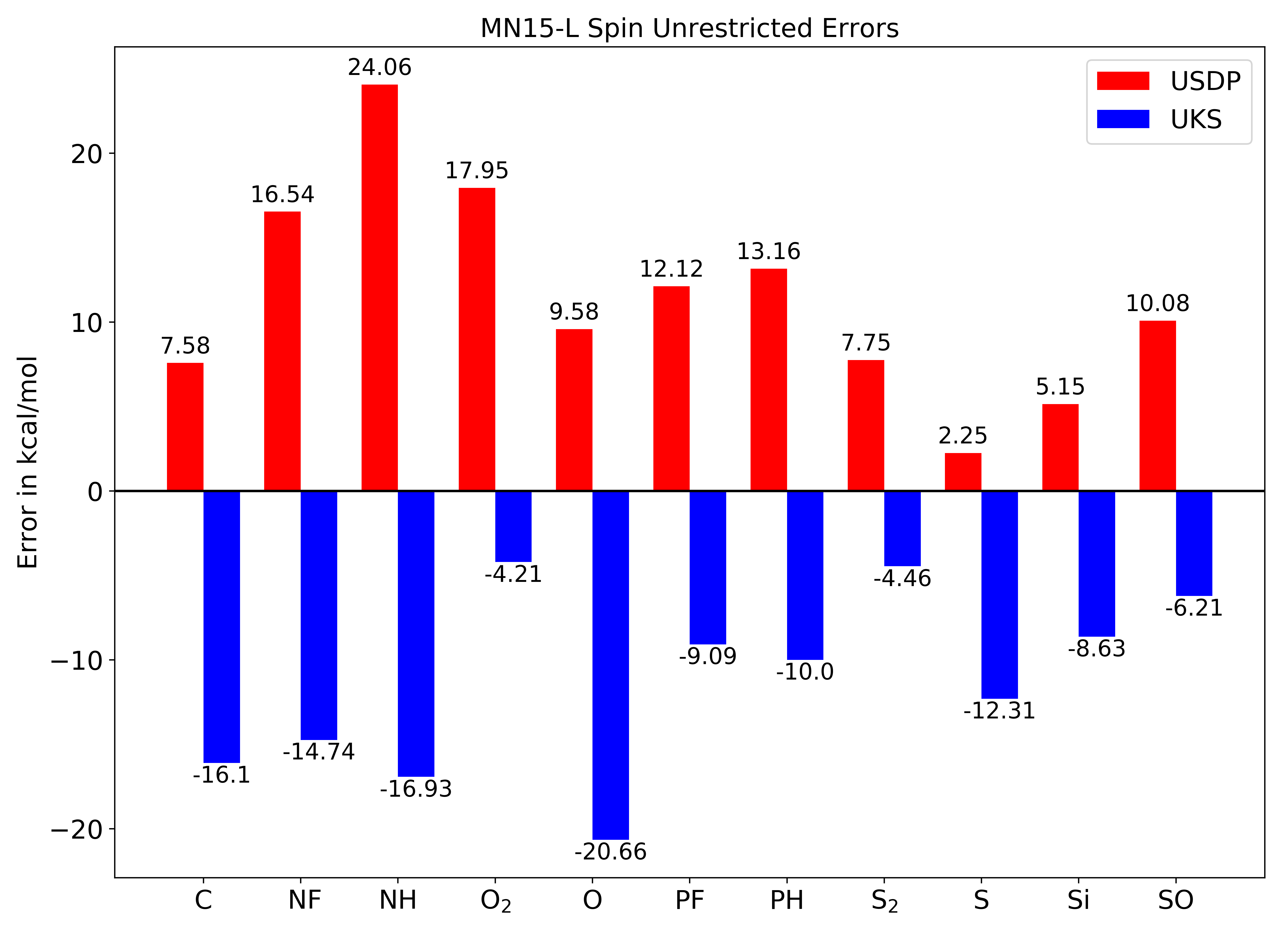}
    \end{minipage}
    \caption{Errors in kcal/mol with respect to experimental values for each species in our test set with the PBE and MN15-L functionals in traditional KS and SDP DFT. Left Column: spin restricted calculations; Right Column: spin unrestricted calculations.}
    \label{fig:ERRORS}
\end{figure*}

To further analyze the performance of the SDP algorithm, we consider the errors of the singlet-triplet gaps $\Delta E_{\text{ST}}$ of the best performing functional, the GGA functional PBE, and the non-improving MN15-L meta-GGA functional. For each species in our test set the errors are shown in Fig. \ref{fig:ERRORS} for both spin restricted and unrestricted implementations. PBE shows a systematic improvement from a massive underestimation or overestimation of the singlet-triplet gap in UKS-DFT and RKS-DFT respectively. The oxygen and sulfur atoms are the only species that, while still showing significant improvements over the KS-DFT, display a $\Delta E_{\text{ST}}$ from SDP-DFT that is significantly lower than the experimental value. Inspection of the changes in the individual energy components of the PBE functional from a KS to SDP implementation shows particularly large stabilization of the singlets for O and S ($\Delta E$'s of -27.78 kcal/mol and  -25.41 kcal/mol, respectively, compared to an average change of -17.31 kcal/mol) and disproportionate destabilization of the triplet ($\Delta E$'s of 5.76 kcal/mol and 1.80 kcal/mol, respectively, compared to an average change of 0.99 kcal/mol) leading to an underestimation of the singlet-triplet gap. In contrast to PBE, MN15-L in USDP-DFT, which again consistently underestimates the gap in UKS-DFT and overestimates the gap in RKS-DFT for every species, uniformly overestimates $\Delta E_{\text{ST}}$ with large magnitudes and no major outliers. The O and S atoms again present negative deviations from the mean error; however, due to the large, general overestimation of the gap these species now profit from favourable error cancellations.\\

\textit{Conclusions}-We present a new SCF minimization procedure for DFT functionals based on semidefinite programming that allows for inclusion of some strong correlation effects via a non-idempotent 1-electron density in systems exhibiting orbital degeneracies. The SDP-DFT method delivers significant improvements over traditional DFT and, unlike traditional KS-DFT, yields results that are consistent across both spin restricted and spin unrestricted implementations. However, improvements are strongly functional dependent. LDA and GGA functionals show consistent refinements from the SDP procedure while improvements from newer, highly parameterized meta-GGA functionals are inconsistent and minor. In particular the MN15-L functional performs worse in the SDP implementation. \\

The present results are comparable to those achieved by Head-Gordon and co-workers with the use of complex spin-restricted orbitals (presented in reference ~\cite{MHG}). The only modifications in the present method from traditional DFT are based on its most fundamental quantity, the electron density, which through the use of SDP rather than KS-SCF minimization, is allowed to derive from non-idempotent and correlated 1-RDMs. The variation of the tested functionals' response to this change in density reveals flaws in the path of modern functional development. Medvedev and co-workers found in their 2017 paper ~\cite{medvedev} that as functional development has progressed from LDA and GGA to modern, highly parameterized meta- and hyper-GGA functionals their predictions improve although DFT's fundamental quantity, the 1-electron density, has strayed further from the true solution. Our results lead to a complementary conclusion, namely that as we progress up ``Jacob's Ladder" a functional's prediction of electronic properties, in our case the singlet-triplet gap of selected simple open shell systems, exhibits less improvement from a refined electron density, pointing towards systematic overfitting in modern hybrid functionals. Nonetheless, the results show that this problem is not universal to DFT as promising improvements are possible with the simplest functionals. Developments focusing on improving implementations of DFT to yield more accurate electron densities may be a more viable path forward than functional development that is based on prediction-driven parametric fitting. This work presents important first steps towards the use of SDP to resolve strong correlation in a DFT framework through the use of improved densities.

\textit{Acknowledgements}
D.A.M. gratefully acknowledges the U.S. National Science Foundation Grant No. CHE-1565638 and the U.S. Army Research Office (ARO) Grant No. W911NF-16-1-0152.

\bibliography{SDP-DFT-V2}

\begin{thebibliography}{62}%
\makeatletter
\providecommand \@ifxundefined [1]{%
 \@ifx{#1\undefined}
}%
\providecommand \@ifnum [1]{%
 \ifnum #1\expandafter \@firstoftwo
 \else \expandafter \@secondoftwo
 \fi
}%
\providecommand \@ifx [1]{%
 \ifx #1\expandafter \@firstoftwo
 \else \expandafter \@secondoftwo
 \fi
}%
\providecommand \natexlab [1]{#1}%
\providecommand \enquote  [1]{``#1''}%
\providecommand \bibnamefont  [1]{#1}%
\providecommand \bibfnamefont [1]{#1}%
\providecommand \citenamefont [1]{#1}%
\providecommand \href@noop [0]{\@secondoftwo}%
\providecommand \href [0]{\begingroup \@sanitize@url \@href}%
\providecommand \@href[1]{\@@startlink{#1}\@@href}%
\providecommand \@@href[1]{\endgroup#1\@@endlink}%
\providecommand \@sanitize@url [0]{\catcode `\\12\catcode `\$12\catcode
  `\&12\catcode `\#12\catcode `\^12\catcode `\_12\catcode `\%12\relax}%
\providecommand \@@startlink[1]{}%
\providecommand \@@endlink[0]{}%
\providecommand \url  [0]{\begingroup\@sanitize@url \@url }%
\providecommand \@url [1]{\endgroup\@href {#1}{\urlprefix }}%
\providecommand \urlprefix  [0]{URL }%
\providecommand \Eprint [0]{\href }%
\providecommand \doibase [0]{http://dx.doi.org/}%
\providecommand \selectlanguage [0]{\@gobble}%
\providecommand \bibinfo  [0]{\@secondoftwo}%
\providecommand \bibfield  [0]{\@secondoftwo}%
\providecommand \translation [1]{[#1]}%
\providecommand \BibitemOpen [0]{}%
\providecommand \bibitemStop [0]{}%
\providecommand \bibitemNoStop [0]{.\EOS\space}%
\providecommand \EOS [0]{\spacefactor3000\relax}%
\providecommand \BibitemShut  [1]{\csname bibitem#1\endcsname}%
\let\auto@bib@innerbib\@empty
\bibitem [{\citenamefont {Burke}(2012)}]{citationstats}%
  \BibitemOpen
  \bibfield  {author} {\bibinfo {author} {\bibfnamefont {K.}~\bibnamefont
  {Burke}},\ }\href@noop {} {\bibfield  {journal} {\bibinfo  {journal} {J.
  Chem. Phys.}\ }\textbf {\bibinfo {volume} {136}},\ \bibinfo {pages} {150901}
  (\bibinfo {year} {2012})}\BibitemShut {NoStop}%
\bibitem [{\citenamefont {Cohen}\ \emph
  {et~al.}(2008{\natexlab{a}})\citenamefont {Cohen}, \citenamefont
  {Mori-S{\'a}nchez},\ and\ \citenamefont {Yang}}]{failures}%
  \BibitemOpen
  \bibfield  {author} {\bibinfo {author} {\bibfnamefont {A.~J.}\ \bibnamefont
  {Cohen}}, \bibinfo {author} {\bibfnamefont {P.}~\bibnamefont
  {Mori-S{\'a}nchez}}, \ and\ \bibinfo {author} {\bibfnamefont
  {W.}~\bibnamefont {Yang}},\ }\href@noop {} {\bibfield  {journal} {\bibinfo
  {journal} {Science}\ }\textbf {\bibinfo {volume} {321}},\ \bibinfo {pages}
  {792} (\bibinfo {year} {2008}{\natexlab{a}})}\BibitemShut {NoStop}%
\bibitem [{\citenamefont {Grimme}(2011)}]{disp}%
  \BibitemOpen
  \bibfield  {author} {\bibinfo {author} {\bibfnamefont {S.}~\bibnamefont
  {Grimme}},\ }\href@noop {} {\bibfield  {journal} {\bibinfo  {journal} {Wiley
  Interdisciplinary Reviews: Computational Molecular Science}\ }\textbf
  {\bibinfo {volume} {1}},\ \bibinfo {pages} {211} (\bibinfo {year}
  {2011})}\BibitemShut {NoStop}%
\bibitem [{\citenamefont {Polo}\ \emph {et~al.}(2002)\citenamefont {Polo},
  \citenamefont {Kraka},\ and\ \citenamefont {Cremer}}]{SIE}%
  \BibitemOpen
  \bibfield  {author} {\bibinfo {author} {\bibfnamefont {V.}~\bibnamefont
  {Polo}}, \bibinfo {author} {\bibfnamefont {E.}~\bibnamefont {Kraka}}, \ and\
  \bibinfo {author} {\bibfnamefont {D.}~\bibnamefont {Cremer}},\ }\href@noop {}
  {\bibfield  {journal} {\bibinfo  {journal} {Mol. Phys.}\ }\textbf {\bibinfo
  {volume} {100}},\ \bibinfo {pages} {1771} (\bibinfo {year}
  {2002})}\BibitemShut {NoStop}%
\bibitem [{\citenamefont {Perdew}\ and\ \citenamefont {Zunger}(1981)}]{Perdew}%
  \BibitemOpen
  \bibfield  {author} {\bibinfo {author} {\bibfnamefont {J.~P.}\ \bibnamefont
  {Perdew}}\ and\ \bibinfo {author} {\bibfnamefont {A.}~\bibnamefont
  {Zunger}},\ }\href@noop {} {\bibfield  {journal} {\bibinfo  {journal} {Phys.
  Rev. B}\ }\textbf {\bibinfo {volume} {23}},\ \bibinfo {pages} {5048}
  (\bibinfo {year} {1981})}\BibitemShut {NoStop}%
\bibitem [{\citenamefont {Cohen}\ \emph {et~al.}(2012)\citenamefont {Cohen},
  \citenamefont {Mori-Sánchez},\ and\ \citenamefont {Yang}}]{yang}%
  \BibitemOpen
  \bibfield  {author} {\bibinfo {author} {\bibfnamefont {A.~J.}\ \bibnamefont
  {Cohen}}, \bibinfo {author} {\bibfnamefont {P.}~\bibnamefont
  {Mori-Sánchez}}, \ and\ \bibinfo {author} {\bibfnamefont {W.}~\bibnamefont
  {Yang}},\ }\href@noop {} {\bibfield  {journal} {\bibinfo  {journal} {Chem.
  Rev.}\ }\textbf {\bibinfo {volume} {112}},\ \bibinfo {pages} {289} (\bibinfo
  {year} {2012})}\BibitemShut {NoStop}%
\bibitem [{\citenamefont {Cohen}\ \emph
  {et~al.}(2008{\natexlab{b}})\citenamefont {Cohen}, \citenamefont
  {Mori-Sánchez},\ and\ \citenamefont {Yang}}]{constfrac}%
  \BibitemOpen
  \bibfield  {author} {\bibinfo {author} {\bibfnamefont {A.~J.}\ \bibnamefont
  {Cohen}}, \bibinfo {author} {\bibfnamefont {P.}~\bibnamefont
  {Mori-Sánchez}}, \ and\ \bibinfo {author} {\bibfnamefont {W.}~\bibnamefont
  {Yang}},\ }\href@noop {} {\bibfield  {journal} {\bibinfo  {journal} {J. Chem.
  Phys.}\ }\textbf {\bibinfo {volume} {129}},\ \bibinfo {pages} {121104}
  (\bibinfo {year} {2008}{\natexlab{b}})}\BibitemShut {NoStop}%
\bibitem [{\citenamefont {Medvedev}\ \emph {et~al.}(2017)\citenamefont
  {Medvedev}, \citenamefont {Bushmarinov}, \citenamefont {Sun}, \citenamefont
  {Perdew},\ and\ \citenamefont {Lyssenko}}]{medvedev}%
  \BibitemOpen
  \bibfield  {author} {\bibinfo {author} {\bibfnamefont {M.~G.}\ \bibnamefont
  {Medvedev}}, \bibinfo {author} {\bibfnamefont {I.~S.}\ \bibnamefont
  {Bushmarinov}}, \bibinfo {author} {\bibfnamefont {J.}~\bibnamefont {Sun}},
  \bibinfo {author} {\bibfnamefont {J.~P.}\ \bibnamefont {Perdew}}, \ and\
  \bibinfo {author} {\bibfnamefont {K.~A.}\ \bibnamefont {Lyssenko}},\
  }\href@noop {} {\bibfield  {journal} {\bibinfo  {journal} {Science}\ }\textbf
  {\bibinfo {volume} {355}},\ \bibinfo {pages} {49} (\bibinfo {year}
  {2017})}\BibitemShut {NoStop}%
\bibitem [{\citenamefont {Perdew}\ \emph {et~al.}(2005)\citenamefont {Perdew},
  \citenamefont {Ruzsinszky}, \citenamefont {Tao}, \citenamefont {Staroverov},
  \citenamefont {Scuseria},\ and\ \citenamefont {Csonka}}]{fitting}%
  \BibitemOpen
  \bibfield  {author} {\bibinfo {author} {\bibfnamefont {J.~P.}\ \bibnamefont
  {Perdew}}, \bibinfo {author} {\bibfnamefont {A.}~\bibnamefont {Ruzsinszky}},
  \bibinfo {author} {\bibfnamefont {J.}~\bibnamefont {Tao}}, \bibinfo {author}
  {\bibfnamefont {V.~N.}\ \bibnamefont {Staroverov}}, \bibinfo {author}
  {\bibfnamefont {G.~E.}\ \bibnamefont {Scuseria}}, \ and\ \bibinfo {author}
  {\bibfnamefont {G.~I.}\ \bibnamefont {Csonka}},\ }\href@noop {} {\bibfield
  {journal} {\bibinfo  {journal} {J. Chem. Phys.}\ }\textbf {\bibinfo {volume}
  {123}},\ \bibinfo {pages} {062201} (\bibinfo {year} {2005})}\BibitemShut
  {NoStop}%
\bibitem [{\citenamefont {Lee}\ \emph {et~al.}(2019)\citenamefont {Lee},
  \citenamefont {Bertels}, \citenamefont {Small},\ and\ \citenamefont
  {Head-Gordon}}]{MHG}%
  \BibitemOpen
  \bibfield  {author} {\bibinfo {author} {\bibfnamefont {J.}~\bibnamefont
  {Lee}}, \bibinfo {author} {\bibfnamefont {L.~W.}\ \bibnamefont {Bertels}},
  \bibinfo {author} {\bibfnamefont {D.~W.}\ \bibnamefont {Small}}, \ and\
  \bibinfo {author} {\bibfnamefont {M.}~\bibnamefont {Head-Gordon}},\
  }\href@noop {} {\bibfield  {journal} {\bibinfo  {journal} {Phys. Rev. Lett.}\
  }\textbf {\bibinfo {volume} {123}},\ \bibinfo {pages} {113001} (\bibinfo
  {year} {2019})}\BibitemShut {NoStop}%
\bibitem [{\citenamefont {Mazziotti}(2011)}]{SDP2}%
  \BibitemOpen
  \bibfield  {author} {\bibinfo {author} {\bibfnamefont {D.~A.}\ \bibnamefont
  {Mazziotti}},\ }\href@noop {} {\bibfield  {journal} {\bibinfo  {journal}
  {Phys. Rev. Lett.}\ }\textbf {\bibinfo {volume} {106}},\ \bibinfo {pages}
  {083001} (\bibinfo {year} {2011})}\BibitemShut {NoStop}%
\bibitem [{\citenamefont {Mazziotti}(2004)}]{SDP1}%
  \BibitemOpen
  \bibfield  {author} {\bibinfo {author} {\bibfnamefont {D.~A.}\ \bibnamefont
  {Mazziotti}},\ }\href@noop {} {\bibfield  {journal} {\bibinfo  {journal}
  {Phys. Rev. Lett.}\ }\textbf {\bibinfo {volume} {93}},\ \bibinfo {pages}
  {213001} (\bibinfo {year} {2004})}\BibitemShut {NoStop}%
\bibitem [{\citenamefont {Montgomery}\ and\ \citenamefont
  {Mazziotti}(2018)}]{V22}%
  \BibitemOpen
  \bibfield  {author} {\bibinfo {author} {\bibfnamefont {J.~M.}\ \bibnamefont
  {Montgomery}}\ and\ \bibinfo {author} {\bibfnamefont {D.~A.}\ \bibnamefont
  {Mazziotti}},\ }\href@noop {} {\bibfield  {journal} {\bibinfo  {journal} {J.
  Phys. Chem. A}\ }\textbf {\bibinfo {volume} {122}},\ \bibinfo {pages} {4988}
  (\bibinfo {year} {2018})}\BibitemShut {NoStop}%
\bibitem [{\citenamefont {Xie}\ \emph {et~al.}(2020)\citenamefont {Xie},
  \citenamefont {Boyn}, \citenamefont {Filatov}, \citenamefont {McNeece},
  \citenamefont {Mazziotti},\ and\ \citenamefont {Anderson}}]{V21}%
  \BibitemOpen
  \bibfield  {author} {\bibinfo {author} {\bibfnamefont {J.}~\bibnamefont
  {Xie}}, \bibinfo {author} {\bibfnamefont {J.-N.}\ \bibnamefont {Boyn}},
  \bibinfo {author} {\bibfnamefont {A.~S.}\ \bibnamefont {Filatov}}, \bibinfo
  {author} {\bibfnamefont {A.~J.}\ \bibnamefont {McNeece}}, \bibinfo {author}
  {\bibfnamefont {D.~A.}\ \bibnamefont {Mazziotti}}, \ and\ \bibinfo {author}
  {\bibfnamefont {J.~S.}\ \bibnamefont {Anderson}},\ }\href@noop {} {\bibfield
  {journal} {\bibinfo  {journal} {Chem. Sci.}\ }\textbf {\bibinfo {volume}
  {11}},\ \bibinfo {pages} {1066} (\bibinfo {year} {2020})}\BibitemShut
  {NoStop}%
\bibitem [{\citenamefont {Schlimgen}\ \emph {et~al.}(2016)\citenamefont
  {Schlimgen}, \citenamefont {Heaps},\ and\ \citenamefont
  {Mazziotti}}]{Schlimgen2016}%
  \BibitemOpen
  \bibfield  {author} {\bibinfo {author} {\bibfnamefont {A.~W.}\ \bibnamefont
  {Schlimgen}}, \bibinfo {author} {\bibfnamefont {C.~W.}\ \bibnamefont
  {Heaps}}, \ and\ \bibinfo {author} {\bibfnamefont {D.~A.}\ \bibnamefont
  {Mazziotti}},\ }\href {\doibase 10.1021/acs.jpclett.5b02547} {\bibfield
  {journal} {\bibinfo  {journal} {J. Phys. Chem. Lett.}\ }\textbf {\bibinfo
  {volume} {7}},\ \bibinfo {pages} {627} (\bibinfo {year} {2016})}\BibitemShut
  {NoStop}%
\bibitem [{\citenamefont {Fosso-Tande}\ \emph {et~al.}(2016)\citenamefont
  {Fosso-Tande}, \citenamefont {Nguyen}, \citenamefont {Gidofalvi},\ and\
  \citenamefont {DePrince}}]{Fosso-Tande2016}%
  \BibitemOpen
  \bibfield  {author} {\bibinfo {author} {\bibfnamefont {J.}~\bibnamefont
  {Fosso-Tande}}, \bibinfo {author} {\bibfnamefont {T.-S.}\ \bibnamefont
  {Nguyen}}, \bibinfo {author} {\bibfnamefont {G.}~\bibnamefont {Gidofalvi}}, \
  and\ \bibinfo {author} {\bibfnamefont {A.~E.}\ \bibnamefont {DePrince}},\
  }\href {\doibase 10.1021/acs.jctc.6b00190} {\bibfield  {journal} {\bibinfo
  {journal} {J. Chem. Theory Comput.}\ }\textbf {\bibinfo {volume} {12}},\
  \bibinfo {pages} {2260} (\bibinfo {year} {2016})}\BibitemShut {NoStop}%
\bibitem [{\citenamefont {Mazziotti}(2016)}]{Mazziotti2016}%
  \BibitemOpen
  \bibfield  {author} {\bibinfo {author} {\bibfnamefont {D.~A.}\ \bibnamefont
  {Mazziotti}},\ }\href {\doibase 10.1103/PhysRevLett.117.153001} {\bibfield
  {journal} {\bibinfo  {journal} {Phys. Rev. Lett.}\ }\textbf {\bibinfo
  {volume} {117}},\ \bibinfo {pages} {153001} (\bibinfo {year}
  {2016})}\BibitemShut {NoStop}%
\bibitem [{\citenamefont {Verstichel}\ \emph {et~al.}(2012)\citenamefont
  {Verstichel}, \citenamefont {van Aggelen}, \citenamefont {Poelmans},\ and\
  \citenamefont {Neck}}]{Verstichel2012}%
  \BibitemOpen
  \bibfield  {author} {\bibinfo {author} {\bibfnamefont {B.}~\bibnamefont
  {Verstichel}}, \bibinfo {author} {\bibfnamefont {H.}~\bibnamefont {van
  Aggelen}}, \bibinfo {author} {\bibfnamefont {W.}~\bibnamefont {Poelmans}}, \
  and\ \bibinfo {author} {\bibfnamefont {D.~V.}\ \bibnamefont {Neck}},\ }\href
  {\doibase 10.1103/physrevlett.108.213001} {\bibfield  {journal} {\bibinfo
  {journal} {Phys. Rev. Lett.}\ }\textbf {\bibinfo {volume} {108}} (\bibinfo
  {year} {2012}),\ 10.1103/physrevlett.108.213001}\BibitemShut {NoStop}%
\bibitem [{\citenamefont {Shenvi}\ and\ \citenamefont
  {Izmaylov}(2010)}]{Shenvi2010}%
  \BibitemOpen
  \bibfield  {author} {\bibinfo {author} {\bibfnamefont {N.}~\bibnamefont
  {Shenvi}}\ and\ \bibinfo {author} {\bibfnamefont {A.~F.}\ \bibnamefont
  {Izmaylov}},\ }\href {\doibase 10.1103/physrevlett.105.213003} {\bibfield
  {journal} {\bibinfo  {journal} {Phys. Rev. Lett.}\ }\textbf {\bibinfo
  {volume} {105}} (\bibinfo {year} {2010}),\
  10.1103/physrevlett.105.213003}\BibitemShut {NoStop}%
\bibitem [{\citenamefont {Canc{\`{e}}s}\ \emph {et~al.}(2006)\citenamefont
  {Canc{\`{e}}s}, \citenamefont {Stoltz},\ and\ \citenamefont
  {Lewin}}]{Cances2006}%
  \BibitemOpen
  \bibfield  {author} {\bibinfo {author} {\bibfnamefont {E.}~\bibnamefont
  {Canc{\`{e}}s}}, \bibinfo {author} {\bibfnamefont {G.}~\bibnamefont
  {Stoltz}}, \ and\ \bibinfo {author} {\bibfnamefont {M.}~\bibnamefont
  {Lewin}},\ }\href {\doibase 10.1063/1.2222358} {\bibfield  {journal}
  {\bibinfo  {journal} {J. Chem. Phys.}\ }\textbf {\bibinfo {volume} {125}},\
  \bibinfo {pages} {064101} (\bibinfo {year} {2006})}\BibitemShut {NoStop}%
\bibitem [{\citenamefont {Zhao}\ \emph {et~al.}(2004)\citenamefont {Zhao},
  \citenamefont {Braams}, \citenamefont {Fukuda}, \citenamefont {Overton},\
  and\ \citenamefont {Percus}}]{Zhao2004}%
  \BibitemOpen
  \bibfield  {author} {\bibinfo {author} {\bibfnamefont {Z.}~\bibnamefont
  {Zhao}}, \bibinfo {author} {\bibfnamefont {B.~J.}\ \bibnamefont {Braams}},
  \bibinfo {author} {\bibfnamefont {M.}~\bibnamefont {Fukuda}}, \bibinfo
  {author} {\bibfnamefont {M.~L.}\ \bibnamefont {Overton}}, \ and\ \bibinfo
  {author} {\bibfnamefont {J.~K.}\ \bibnamefont {Percus}},\ }\href {\doibase
  10.1063/1.1636721} {\bibfield  {journal} {\bibinfo  {journal} {J. Chem.
  Phys.}\ }\textbf {\bibinfo {volume} {120}},\ \bibinfo {pages} {2095}
  (\bibinfo {year} {2004})}\BibitemShut {NoStop}%
\bibitem [{\citenamefont {Mazziotti}(2002)}]{Mazziotti2002b}%
  \BibitemOpen
  \bibfield  {author} {\bibinfo {author} {\bibfnamefont {D.~A.}\ \bibnamefont
  {Mazziotti}},\ }\href {\doibase 10.1103/PhysRevA.65.062511} {\bibfield
  {journal} {\bibinfo  {journal} {Phys. Rev. A}\ }\textbf {\bibinfo {volume}
  {65}},\ \bibinfo {pages} {062511} (\bibinfo {year} {2002})}\BibitemShut
  {NoStop}%
\bibitem [{\citenamefont {Nakata}\ \emph {et~al.}(2001)\citenamefont {Nakata},
  \citenamefont {Nakatsuji}, \citenamefont {Ehara}, \citenamefont {Fukuda},
  \citenamefont {Nakata},\ and\ \citenamefont {Fujisawa}}]{Nakata2001}%
  \BibitemOpen
  \bibfield  {author} {\bibinfo {author} {\bibfnamefont {M.}~\bibnamefont
  {Nakata}}, \bibinfo {author} {\bibfnamefont {H.}~\bibnamefont {Nakatsuji}},
  \bibinfo {author} {\bibfnamefont {M.}~\bibnamefont {Ehara}}, \bibinfo
  {author} {\bibfnamefont {M.}~\bibnamefont {Fukuda}}, \bibinfo {author}
  {\bibfnamefont {K.}~\bibnamefont {Nakata}}, \ and\ \bibinfo {author}
  {\bibfnamefont {K.}~\bibnamefont {Fujisawa}},\ }\href {\doibase
  10.1063/1.1360199} {\bibfield  {journal} {\bibinfo  {journal} {J. Chem.
  Phys.}\ }\textbf {\bibinfo {volume} {114}},\ \bibinfo {pages} {8282}
  (\bibinfo {year} {2001})}\BibitemShut {NoStop}%
\bibitem [{\citenamefont {Garrod}\ and\ \citenamefont
  {Percus}(1964)}]{Garrod1964}%
  \BibitemOpen
  \bibfield  {author} {\bibinfo {author} {\bibfnamefont {C.}~\bibnamefont
  {Garrod}}\ and\ \bibinfo {author} {\bibfnamefont {J.~K.}\ \bibnamefont
  {Percus}},\ }\href {\doibase 10.1063/1.1704098} {\bibfield  {journal}
  {\bibinfo  {journal} {J. Math. Phys.}\ }\textbf {\bibinfo {volume} {5}},\
  \bibinfo {pages} {1756} (\bibinfo {year} {1964})}\BibitemShut {NoStop}%
\bibitem [{\citenamefont {Coleman}(1963)}]{Coleman1963}%
  \BibitemOpen
  \bibfield  {author} {\bibinfo {author} {\bibfnamefont {A.~J.}\ \bibnamefont
  {Coleman}},\ }\href {\doibase 10.1103/revmodphys.35.668} {\bibfield
  {journal} {\bibinfo  {journal} {Rev. Mod. Phys.}\ }\textbf {\bibinfo {volume}
  {35}},\ \bibinfo {pages} {668} (\bibinfo {year} {1963})}\BibitemShut
  {NoStop}%
\bibitem [{\citenamefont {Coleman}(2007)}]{Coleman2007}%
  \BibitemOpen
  \bibfield  {author} {\bibinfo {author} {\bibfnamefont {A.~J.}\ \bibnamefont
  {Coleman}},\ }in\ \href {\doibase 10.1002/9780470106600.ch1} {\emph {\bibinfo
  {booktitle} {Reduced-Density-Matrix Mechanics: With Application to
  Many-Electron Atoms and Molecules}}}\ (\bibinfo  {publisher} {John Wiley {\&}
  Sons, Inc.},\ \bibinfo {year} {2007})\ pp.\ \bibinfo {pages}
  {1--9}\BibitemShut {NoStop}%
\bibitem [{\citenamefont {Mazziotti}(2012)}]{Mazziotti2012b}%
  \BibitemOpen
  \bibfield  {author} {\bibinfo {author} {\bibfnamefont {D.~A.}\ \bibnamefont
  {Mazziotti}},\ }\href {\doibase 10.1103/PhysRevLett.108.263002} {\bibfield
  {journal} {\bibinfo  {journal} {Phys. Rev. Lett.}\ }\textbf {\bibinfo
  {volume} {108}},\ \bibinfo {pages} {263002} (\bibinfo {year}
  {2012})}\BibitemShut {NoStop}%
\bibitem [{\citenamefont {Schmidt}\ \emph {et~al.}(2019)\citenamefont
  {Schmidt}, \citenamefont {Benavides-Riveros},\ and\ \citenamefont
  {Marques}}]{Schmidt2019}%
  \BibitemOpen
  \bibfield  {author} {\bibinfo {author} {\bibfnamefont {J.}~\bibnamefont
  {Schmidt}}, \bibinfo {author} {\bibfnamefont {C.~L.}\ \bibnamefont
  {Benavides-Riveros}}, \ and\ \bibinfo {author} {\bibfnamefont {M.~A.~L.}\
  \bibnamefont {Marques}},\ }\href {\doibase 10.1103/PhysRevB.99.224502}
  {\bibfield  {journal} {\bibinfo  {journal} {Phys. Rev. B}\ }\textbf {\bibinfo
  {volume} {99}},\ \bibinfo {pages} {224502} (\bibinfo {year}
  {2019})}\BibitemShut {NoStop}%
\bibitem [{\citenamefont {Schilling}\ and\ \citenamefont
  {Schilling}(2019)}]{Schilling2019}%
  \BibitemOpen
  \bibfield  {author} {\bibinfo {author} {\bibfnamefont {C.}~\bibnamefont
  {Schilling}}\ and\ \bibinfo {author} {\bibfnamefont {R.}~\bibnamefont
  {Schilling}},\ }\href {\doibase 10.1103/PhysRevLett.122.013001} {\bibfield
  {journal} {\bibinfo  {journal} {Phys. Rev. Lett.}\ }\textbf {\bibinfo
  {volume} {122}},\ \bibinfo {pages} {013001} (\bibinfo {year}
  {2019})}\BibitemShut {NoStop}%
\bibitem [{\citenamefont {Piris}(2018)}]{Piris2018}%
  \BibitemOpen
  \bibfield  {author} {\bibinfo {author} {\bibfnamefont {M.}~\bibnamefont
  {Piris}},\ }\href {\doibase 10.1103/PhysRevA.98.022504} {\bibfield  {journal}
  {\bibinfo  {journal} {Phys. Rev. A}\ }\textbf {\bibinfo {volume} {98}},\
  \bibinfo {pages} {022504} (\bibinfo {year} {2018})}\BibitemShut {NoStop}%
\bibitem [{\citenamefont {Piris}(2017)}]{Piris2017a}%
  \BibitemOpen
  \bibfield  {author} {\bibinfo {author} {\bibfnamefont {M.}~\bibnamefont
  {Piris}},\ }\href {\doibase 10.1103/PhysRevLett.119.063002} {\bibfield
  {journal} {\bibinfo  {journal} {Phys. Rev. Lett.}\ }\textbf {\bibinfo
  {volume} {119}},\ \bibinfo {pages} {063002} (\bibinfo {year}
  {2017})}\BibitemShut {NoStop}%
\bibitem [{\citenamefont {Sharma}\ \emph {et~al.}(2013)\citenamefont {Sharma},
  \citenamefont {Dewhurst}, \citenamefont {Shallcross},\ and\ \citenamefont
  {Gross}}]{Sharma2013}%
  \BibitemOpen
  \bibfield  {author} {\bibinfo {author} {\bibfnamefont {S.}~\bibnamefont
  {Sharma}}, \bibinfo {author} {\bibfnamefont {J.~K.}\ \bibnamefont
  {Dewhurst}}, \bibinfo {author} {\bibfnamefont {S.}~\bibnamefont
  {Shallcross}}, \ and\ \bibinfo {author} {\bibfnamefont {E.~K.~U.}\
  \bibnamefont {Gross}},\ }\href {\doibase 10.1103/physrevlett.110.116403}
  {\bibfield  {journal} {\bibinfo  {journal} {Phys. Rev. Lett.}\ }\textbf
  {\bibinfo {volume} {110}} (\bibinfo {year} {2013}),\
  10.1103/physrevlett.110.116403}\BibitemShut {NoStop}%
\bibitem [{\citenamefont {Rohr}\ \emph {et~al.}(2008)\citenamefont {Rohr},
  \citenamefont {Pernal}, \citenamefont {Gritsenko},\ and\ \citenamefont
  {Baerends}}]{Rohr2008}%
  \BibitemOpen
  \bibfield  {author} {\bibinfo {author} {\bibfnamefont {D.~R.}\ \bibnamefont
  {Rohr}}, \bibinfo {author} {\bibfnamefont {K.}~\bibnamefont {Pernal}},
  \bibinfo {author} {\bibfnamefont {O.~V.}\ \bibnamefont {Gritsenko}}, \ and\
  \bibinfo {author} {\bibfnamefont {E.~J.}\ \bibnamefont {Baerends}},\ }\href
  {\doibase 10.1063/1.2998201} {\bibfield  {journal} {\bibinfo  {journal} {J.
  Chem. Phys.}\ }\textbf {\bibinfo {volume} {129}},\ \bibinfo {pages} {164105}
  (\bibinfo {year} {2008})}\BibitemShut {NoStop}%
\bibitem [{\citenamefont {Mazziotti}(2001)}]{Mazziotti2001b}%
  \BibitemOpen
  \bibfield  {author} {\bibinfo {author} {\bibfnamefont {D.~A.}\ \bibnamefont
  {Mazziotti}},\ }\href@noop {} {\bibfield  {journal} {\bibinfo  {journal}
  {Chem. Phys. Lett.}\ } (\bibinfo {year} {2001})}\BibitemShut {NoStop}%
\bibitem [{\citenamefont {Goedecker}\ and\ \citenamefont
  {Umrigar}(1998)}]{Goedecker1998}%
  \BibitemOpen
  \bibfield  {author} {\bibinfo {author} {\bibfnamefont {S.}~\bibnamefont
  {Goedecker}}\ and\ \bibinfo {author} {\bibfnamefont {C.~J.}\ \bibnamefont
  {Umrigar}},\ }\href {\doibase 10.1103/physrevlett.81.866} {\bibfield
  {journal} {\bibinfo  {journal} {Phys. Rev. Lett.}\ }\textbf {\bibinfo
  {volume} {81}},\ \bibinfo {pages} {866} (\bibinfo {year} {1998})}\BibitemShut
  {NoStop}%
\bibitem [{\citenamefont {Buijse}\ and\ \citenamefont
  {Baerends}(2002)}]{BUIJSE2002}%
  \BibitemOpen
  \bibfield  {author} {\bibinfo {author} {\bibfnamefont {M.~A.}\ \bibnamefont
  {Buijse}}\ and\ \bibinfo {author} {\bibfnamefont {E.~J.}\ \bibnamefont
  {Baerends}},\ }\href {\doibase 10.1080/00268970110070243} {\bibfield
  {journal} {\bibinfo  {journal} {Mol. Phys.}\ }\textbf {\bibinfo {volume}
  {100}},\ \bibinfo {pages} {401} (\bibinfo {year} {2002})}\BibitemShut
  {NoStop}%
\bibitem [{\citenamefont {Miehlich}\ \emph {et~al.}(1997)\citenamefont
  {Miehlich}, \citenamefont {Stoll},\ and\ \citenamefont {Savin}}]{MCDFT1}%
  \BibitemOpen
  \bibfield  {author} {\bibinfo {author} {\bibfnamefont {B.}~\bibnamefont
  {Miehlich}}, \bibinfo {author} {\bibfnamefont {H.}~\bibnamefont {Stoll}}, \
  and\ \bibinfo {author} {\bibfnamefont {A.}~\bibnamefont {Savin}},\
  }\href@noop {} {\bibfield  {journal} {\bibinfo  {journal} {Mol. Phys.}\
  }\textbf {\bibinfo {volume} {91}},\ \bibinfo {pages} {527} (\bibinfo {year}
  {1997})}\BibitemShut {NoStop}%
\bibitem [{\citenamefont {Kurzweil}\ \emph {et~al.}(2009)\citenamefont
  {Kurzweil}, \citenamefont {Lawler},\ and\ \citenamefont
  {Head-Gordon}}]{MCDFT2}%
  \BibitemOpen
  \bibfield  {author} {\bibinfo {author} {\bibfnamefont {Y.}~\bibnamefont
  {Kurzweil}}, \bibinfo {author} {\bibfnamefont {K.~V.}\ \bibnamefont
  {Lawler}}, \ and\ \bibinfo {author} {\bibfnamefont {M.}~\bibnamefont
  {Head-Gordon}},\ }\href@noop {} {\bibfield  {journal} {\bibinfo  {journal}
  {Mol. Phys.}\ }\textbf {\bibinfo {volume} {107}},\ \bibinfo {pages} {2103}
  (\bibinfo {year} {2009})}\BibitemShut {NoStop}%
\bibitem [{\citenamefont {Li~Manni}\ \emph {et~al.}(2014)\citenamefont
  {Li~Manni}, \citenamefont {Carlson}, \citenamefont {Luo}, \citenamefont {Ma},
  \citenamefont {Olsen}, \citenamefont {Truhlar},\ and\ \citenamefont
  {Gagliardi}}]{MCPDFT1}%
  \BibitemOpen
  \bibfield  {author} {\bibinfo {author} {\bibfnamefont {G.}~\bibnamefont
  {Li~Manni}}, \bibinfo {author} {\bibfnamefont {R.~K.}\ \bibnamefont
  {Carlson}}, \bibinfo {author} {\bibfnamefont {S.}~\bibnamefont {Luo}},
  \bibinfo {author} {\bibfnamefont {D.}~\bibnamefont {Ma}}, \bibinfo {author}
  {\bibfnamefont {J.}~\bibnamefont {Olsen}}, \bibinfo {author} {\bibfnamefont
  {D.~G.}\ \bibnamefont {Truhlar}}, \ and\ \bibinfo {author} {\bibfnamefont
  {L.}~\bibnamefont {Gagliardi}},\ }\href@noop {} {\bibfield  {journal}
  {\bibinfo  {journal} {J. Chem. Theory Comput.}\ }\textbf {\bibinfo {volume}
  {10}},\ \bibinfo {pages} {3669} (\bibinfo {year} {2014})}\BibitemShut
  {NoStop}%
\bibitem [{\citenamefont {Bao}\ \emph {et~al.}(2016)\citenamefont {Bao},
  \citenamefont {Sand}, \citenamefont {Gagliardi},\ and\ \citenamefont
  {Truhlar}}]{MCPDFT2}%
  \BibitemOpen
  \bibfield  {author} {\bibinfo {author} {\bibfnamefont {J.~L.}\ \bibnamefont
  {Bao}}, \bibinfo {author} {\bibfnamefont {A.}~\bibnamefont {Sand}}, \bibinfo
  {author} {\bibfnamefont {L.}~\bibnamefont {Gagliardi}}, \ and\ \bibinfo
  {author} {\bibfnamefont {D.~G.}\ \bibnamefont {Truhlar}},\ }\href@noop {}
  {\bibfield  {journal} {\bibinfo  {journal} {J. Chem. Theory Comput.}\
  }\textbf {\bibinfo {volume} {12}},\ \bibinfo {pages} {4274} (\bibinfo {year}
  {2016})}\BibitemShut {NoStop}%
\bibitem [{\citenamefont {Kohn}\ and\ \citenamefont
  {Sham}(1965{\natexlab{a}})}]{KSDFT}%
  \BibitemOpen
  \bibfield  {author} {\bibinfo {author} {\bibfnamefont {W.}~\bibnamefont
  {Kohn}}\ and\ \bibinfo {author} {\bibfnamefont {L.~J.}\ \bibnamefont
  {Sham}},\ }\href@noop {} {\bibfield  {journal} {\bibinfo  {journal} {Phys.
  Rev.}\ }\textbf {\bibinfo {volume} {140}},\ \bibinfo {pages} {A1133}
  (\bibinfo {year} {1965}{\natexlab{a}})}\BibitemShut {NoStop}%
\bibitem [{\citenamefont {Veeraraghavan}\ and\ \citenamefont
  {Mazziotti}(2014{\natexlab{a}})}]{SDPHF1}%
  \BibitemOpen
  \bibfield  {author} {\bibinfo {author} {\bibfnamefont {S.}~\bibnamefont
  {Veeraraghavan}}\ and\ \bibinfo {author} {\bibfnamefont {D.~A.}\ \bibnamefont
  {Mazziotti}},\ }\href@noop {} {\bibfield  {journal} {\bibinfo  {journal}
  {Phys. Rev. A}\ }\textbf {\bibinfo {volume} {89}},\ \bibinfo {pages} {010502}
  (\bibinfo {year} {2014}{\natexlab{a}})}\BibitemShut {NoStop}%
\bibitem [{\citenamefont {Veeraraghavan}\ and\ \citenamefont
  {Mazziotti}(2014{\natexlab{b}})}]{SDPHF2}%
  \BibitemOpen
  \bibfield  {author} {\bibinfo {author} {\bibfnamefont {S.}~\bibnamefont
  {Veeraraghavan}}\ and\ \bibinfo {author} {\bibfnamefont {D.~A.}\ \bibnamefont
  {Mazziotti}},\ }\href@noop {} {\bibfield  {journal} {\bibinfo  {journal} {J.
  Chem. Phys.}\ }\textbf {\bibinfo {volume} {140}},\ \bibinfo {pages} {124106}
  (\bibinfo {year} {2014}{\natexlab{b}})}\BibitemShut {NoStop}%
\bibitem [{\citenamefont {Veeraraghavan}\ and\ \citenamefont
  {Mazziotti}(2015)}]{Veeraraghavan2015}%
  \BibitemOpen
  \bibfield  {author} {\bibinfo {author} {\bibfnamefont {S.}~\bibnamefont
  {Veeraraghavan}}\ and\ \bibinfo {author} {\bibfnamefont {D.~A.}\ \bibnamefont
  {Mazziotti}},\ }\href {\doibase 10.1103/PhysRevA.92.022512} {\bibfield
  {journal} {\bibinfo  {journal} {Phys. Rev. A}\ }\textbf {\bibinfo {volume}
  {92}},\ \bibinfo {pages} {022512} (\bibinfo {year} {2015})}\BibitemShut
  {NoStop}%
\bibitem [{\citenamefont {Vandenberghe}\ and\ \citenamefont
  {Boyd}(1996)}]{SDP3}%
  \BibitemOpen
  \bibfield  {author} {\bibinfo {author} {\bibfnamefont {L.}~\bibnamefont
  {Vandenberghe}}\ and\ \bibinfo {author} {\bibfnamefont {S.}~\bibnamefont
  {Boyd}},\ }\href@noop {} {\bibfield  {journal} {\bibinfo  {journal} {SIAM
  Rev.}\ }\textbf {\bibinfo {volume} {38}},\ \bibinfo {pages} {49} (\bibinfo
  {year} {1996})}\BibitemShut {NoStop}%
\bibitem [{\citenamefont {Erdahl}(1979)}]{SDP4}%
  \BibitemOpen
  \bibfield  {author} {\bibinfo {author} {\bibfnamefont {R.}~\bibnamefont
  {Erdahl}},\ }\href@noop {} {\bibfield  {journal} {\bibinfo  {journal} {Rep.
  Math. Phys.}\ }\textbf {\bibinfo {volume} {15}},\ \bibinfo {pages} {147 }
  (\bibinfo {year} {1979})}\BibitemShut {NoStop}%
\bibitem [{\citenamefont {Perdew}\ and\ \citenamefont
  {Schmidt}(2001)}]{Ladder}%
  \BibitemOpen
  \bibfield  {author} {\bibinfo {author} {\bibfnamefont {J.~P.}\ \bibnamefont
  {Perdew}}\ and\ \bibinfo {author} {\bibfnamefont {K.}~\bibnamefont
  {Schmidt}},\ }\href@noop {} {\bibfield  {journal} {\bibinfo  {journal} {AIP
  Conference Proceedings}\ }\textbf {\bibinfo {volume} {577}},\ \bibinfo
  {pages} {1} (\bibinfo {year} {2001})}\BibitemShut {NoStop}%
\bibitem [{\citenamefont {Hohenberg}\ and\ \citenamefont {Kohn}(1964)}]{LDA1}%
  \BibitemOpen
  \bibfield  {author} {\bibinfo {author} {\bibfnamefont {P.}~\bibnamefont
  {Hohenberg}}\ and\ \bibinfo {author} {\bibfnamefont {W.}~\bibnamefont
  {Kohn}},\ }\href@noop {} {\bibfield  {journal} {\bibinfo  {journal} {Phys.
  Rev.}\ }\textbf {\bibinfo {volume} {136}},\ \bibinfo {pages} {B864} (\bibinfo
  {year} {1964})}\BibitemShut {NoStop}%
\bibitem [{\citenamefont {Kohn}\ and\ \citenamefont
  {Sham}(1965{\natexlab{b}})}]{LDA2}%
  \BibitemOpen
  \bibfield  {author} {\bibinfo {author} {\bibfnamefont {W.}~\bibnamefont
  {Kohn}}\ and\ \bibinfo {author} {\bibfnamefont {L.~J.}\ \bibnamefont
  {Sham}},\ }\href@noop {} {\bibfield  {journal} {\bibinfo  {journal} {Phys.
  Rev.}\ }\textbf {\bibinfo {volume} {140}},\ \bibinfo {pages} {A1133}
  (\bibinfo {year} {1965}{\natexlab{b}})}\BibitemShut {NoStop}%
\bibitem [{\citenamefont {Slater}(1974)}]{Slater}%
  \BibitemOpen
  \bibfield  {author} {\bibinfo {author} {\bibfnamefont {J.~C.}\ \bibnamefont
  {Slater}},\ }\href@noop {} {\bibfield  {journal} {\bibinfo  {journal}
  {Quantum Theory of Molecular and Solids}\ }\textbf {\bibinfo {volume} {4}}
  (\bibinfo {year} {1974})}\BibitemShut {NoStop}%
\bibitem [{\citenamefont {Vosko}\ \emph {et~al.}(1980)\citenamefont {Vosko},
  \citenamefont {Wilk},\ and\ \citenamefont {Nusair}}]{VWN}%
  \BibitemOpen
  \bibfield  {author} {\bibinfo {author} {\bibfnamefont {S.~H.}\ \bibnamefont
  {Vosko}}, \bibinfo {author} {\bibfnamefont {L.}~\bibnamefont {Wilk}}, \ and\
  \bibinfo {author} {\bibfnamefont {M.}~\bibnamefont {Nusair}},\ }\href@noop {}
  {\bibfield  {journal} {\bibinfo  {journal} {Can. J. Phys}\ }\textbf {\bibinfo
  {volume} {58}},\ \bibinfo {pages} {1200} (\bibinfo {year}
  {1980})}\BibitemShut {NoStop}%
\bibitem [{\citenamefont {Perdew}\ and\ \citenamefont {Wang}(1992)}]{SPW}%
  \BibitemOpen
  \bibfield  {author} {\bibinfo {author} {\bibfnamefont {J.~P.}\ \bibnamefont
  {Perdew}}\ and\ \bibinfo {author} {\bibfnamefont {Y.}~\bibnamefont {Wang}},\
  }\href@noop {} {\bibfield  {journal} {\bibinfo  {journal} {Phys. Rev. B}\
  }\textbf {\bibinfo {volume} {45}},\ \bibinfo {pages} {13244} (\bibinfo {year}
  {1992})}\BibitemShut {NoStop}%
\bibitem [{\citenamefont {Perdew}\ \emph {et~al.}(1996)\citenamefont {Perdew},
  \citenamefont {Burke},\ and\ \citenamefont {Ernzerhof}}]{PBE1}%
  \BibitemOpen
  \bibfield  {author} {\bibinfo {author} {\bibfnamefont {J.~P.}\ \bibnamefont
  {Perdew}}, \bibinfo {author} {\bibfnamefont {K.}~\bibnamefont {Burke}}, \
  and\ \bibinfo {author} {\bibfnamefont {M.}~\bibnamefont {Ernzerhof}},\
  }\href@noop {} {\bibfield  {journal} {\bibinfo  {journal} {Phys. Rev. Lett.}\
  }\textbf {\bibinfo {volume} {77}},\ \bibinfo {pages} {3865} (\bibinfo {year}
  {1996})}\BibitemShut {NoStop}%
\bibitem [{\citenamefont {Perdew}\ \emph {et~al.}(1997)\citenamefont {Perdew},
  \citenamefont {Burke},\ and\ \citenamefont {Ernzerhof}}]{PBE2}%
  \BibitemOpen
  \bibfield  {author} {\bibinfo {author} {\bibfnamefont {J.~P.}\ \bibnamefont
  {Perdew}}, \bibinfo {author} {\bibfnamefont {K.}~\bibnamefont {Burke}}, \
  and\ \bibinfo {author} {\bibfnamefont {M.}~\bibnamefont {Ernzerhof}},\
  }\href@noop {} {\bibfield  {journal} {\bibinfo  {journal} {Phys. Rev. Lett.}\
  }\textbf {\bibinfo {volume} {78}},\ \bibinfo {pages} {1396} (\bibinfo {year}
  {1997})}\BibitemShut {NoStop}%
\bibitem [{\citenamefont {Becke}(1988)}]{BLYP1}%
  \BibitemOpen
  \bibfield  {author} {\bibinfo {author} {\bibfnamefont {A.~D.}\ \bibnamefont
  {Becke}},\ }\href@noop {} {\bibfield  {journal} {\bibinfo  {journal} {Phys.
  Rev. A}\ }\textbf {\bibinfo {volume} {38}},\ \bibinfo {pages} {3098}
  (\bibinfo {year} {1988})}\BibitemShut {NoStop}%
\bibitem [{\citenamefont {Lee}\ \emph {et~al.}(1988)\citenamefont {Lee},
  \citenamefont {Yang},\ and\ \citenamefont {Parr}}]{BLYP2}%
  \BibitemOpen
  \bibfield  {author} {\bibinfo {author} {\bibfnamefont {C.}~\bibnamefont
  {Lee}}, \bibinfo {author} {\bibfnamefont {W.}~\bibnamefont {Yang}}, \ and\
  \bibinfo {author} {\bibfnamefont {R.~G.}\ \bibnamefont {Parr}},\ }\href@noop
  {} {\bibfield  {journal} {\bibinfo  {journal} {Phys. Rev. B}\ }\textbf
  {\bibinfo {volume} {37}},\ \bibinfo {pages} {785} (\bibinfo {year}
  {1988})}\BibitemShut {NoStop}%
\bibitem [{\citenamefont {Tao}\ \emph {et~al.}(2003)\citenamefont {Tao},
  \citenamefont {Perdew}, \citenamefont {Staroverov},\ and\ \citenamefont
  {Scuseria}}]{TPSS}%
  \BibitemOpen
  \bibfield  {author} {\bibinfo {author} {\bibfnamefont {J.}~\bibnamefont
  {Tao}}, \bibinfo {author} {\bibfnamefont {J.~P.}\ \bibnamefont {Perdew}},
  \bibinfo {author} {\bibfnamefont {V.~N.}\ \bibnamefont {Staroverov}}, \ and\
  \bibinfo {author} {\bibfnamefont {G.~E.}\ \bibnamefont {Scuseria}},\
  }\href@noop {} {\bibfield  {journal} {\bibinfo  {journal} {Phys. Rev. Lett.}\
  }\textbf {\bibinfo {volume} {91}},\ \bibinfo {pages} {146401} (\bibinfo
  {year} {2003})}\BibitemShut {NoStop}%
\bibitem [{\citenamefont {Sun}\ \emph {et~al.}(2015)\citenamefont {Sun},
  \citenamefont {Ruzsinszky},\ and\ \citenamefont {Perdew}}]{SCAN}%
  \BibitemOpen
  \bibfield  {author} {\bibinfo {author} {\bibfnamefont {J.}~\bibnamefont
  {Sun}}, \bibinfo {author} {\bibfnamefont {A.}~\bibnamefont {Ruzsinszky}}, \
  and\ \bibinfo {author} {\bibfnamefont {J.~P.}\ \bibnamefont {Perdew}},\
  }\href@noop {} {\bibfield  {journal} {\bibinfo  {journal} {Phys. Rev. Lett.}\
  }\textbf {\bibinfo {volume} {115}},\ \bibinfo {pages} {036402} (\bibinfo
  {year} {2015})}\BibitemShut {NoStop}%
\bibitem [{\citenamefont {Yu}\ \emph {et~al.}(2016)\citenamefont {Yu},
  \citenamefont {He},\ and\ \citenamefont {Truhlar}}]{MN15L}%
  \BibitemOpen
  \bibfield  {author} {\bibinfo {author} {\bibfnamefont {H.~S.}\ \bibnamefont
  {Yu}}, \bibinfo {author} {\bibfnamefont {X.}~\bibnamefont {He}}, \ and\
  \bibinfo {author} {\bibfnamefont {D.~G.}\ \bibnamefont {Truhlar}},\
  }\href@noop {} {\bibfield  {journal} {\bibinfo  {journal} {J. Chem. Theory
  Comput.}\ }\textbf {\bibinfo {volume} {12}},\ \bibinfo {pages} {1280}
  (\bibinfo {year} {2016})}\BibitemShut {NoStop}%
\bibitem [{\citenamefont {Mardirossian}\ and\ \citenamefont
  {Head-Gordon}(2015)}]{B97MV}%
  \BibitemOpen
  \bibfield  {author} {\bibinfo {author} {\bibfnamefont {N.}~\bibnamefont
  {Mardirossian}}\ and\ \bibinfo {author} {\bibfnamefont {M.}~\bibnamefont
  {Head-Gordon}},\ }\href@noop {} {\bibfield  {journal} {\bibinfo  {journal}
  {J. Chem. Phys.}\ }\textbf {\bibinfo {volume} {142}},\ \bibinfo {pages}
  {074111} (\bibinfo {year} {2015})}\BibitemShut {NoStop}%
\bibitem [{\citenamefont {Dunning}(1989)}]{basis1}%
  \BibitemOpen
  \bibfield  {author} {\bibinfo {author} {\bibfnamefont {T.~H.}\ \bibnamefont
  {Dunning}},\ }\href@noop {} {\bibfield  {journal} {\bibinfo  {journal} {J.
  Chem. Phys.}\ }\textbf {\bibinfo {volume} {90}},\ \bibinfo {pages} {1007}
  (\bibinfo {year} {1989})}\BibitemShut {NoStop}%
\bibitem [{\citenamefont {Kendall}\ \emph {et~al.}(1992)\citenamefont
  {Kendall}, \citenamefont {Dunning},\ and\ \citenamefont {Harrison}}]{basis2}%
  \BibitemOpen
  \bibfield  {author} {\bibinfo {author} {\bibfnamefont {R.~A.}\ \bibnamefont
  {Kendall}}, \bibinfo {author} {\bibfnamefont {T.~H.}\ \bibnamefont
  {Dunning}}, \ and\ \bibinfo {author} {\bibfnamefont {R.~J.}\ \bibnamefont
  {Harrison}},\ }\href@noop {} {\bibfield  {journal} {\bibinfo  {journal} {J.
  Chem. Phys.}\ }\textbf {\bibinfo {volume} {96}},\ \bibinfo {pages} {6796}
  (\bibinfo {year} {1992})}\BibitemShut {NoStop}%
\end{thebibliography}%

\end{document}